\documentstyle[12pt]{article}
\input{epsfig} \textheight 21 cm \textwidth 16 cm \oddsidemargin
-0.15 cm
\begin{document}


\title{Expansion of a Bose-Einstein Condensate in the Presence of Disorder }

\author{ Boris Shapiro\\
Department of Physics, Technion-Israel Institute of Technology,
Haifa 32000, Israel}
\maketitle

\begin{abstract}
Expansion of a Bose-Einstein condensate (BEC) is studied, in the
presence of a random potential. The expansion is controlled by a
single parameter, $(\mu\tau_{eff} /\hbar)$, where $\mu$ is the
chemical potential, prior to the release of the BEC from the trap,
and $\tau_{eff}$ is a transport relaxation time which
characterizes the strength of the disorder.  Repulsive
interactions (nonlinearity) facilitate transport and can lead to
diffusive spreading of the condensate which, in the absence of
interactions, would have remained localized in the vicinity of its
initial location.


\end{abstract}

\vskip 4cm

There is a number of recent experiments on the interplay between
nonlinearity and disorder \cite{lye, clement1, fort,  clement2,
shulte, segev}. In \cite{lye, clement1, fort,  clement2} the
expansion of a BEC, in the presence of a random potential, has
been studied. In \cite {segev} the authors study propagation of a
light beam through a dielectric sample, with a random refraction
index, under linear and nonlinear conditions. Under certain
conditions both types of experiments can be described with the
Gross-Pitaevskii equation. For the BEC problem the necessary
conditions are specified, e.g., in \cite {string}. For optical
waves the Gross-Pitaevskii equation emerges within the paraxial
approximation (see \cite{segev} and references therein). The
equation is:
\begin{equation}\label{G1}
 i\hbar \frac{\partial \Psi}{\partial t}=
-\frac{\hbar^2}{2m}\triangle\Psi
 + V({\bf r})\Psi +g |\Psi|^2 \Psi,
\end{equation}
where $m$ is the atom mass, $g = 4\pi a \hbar^2 /m$ is the
interaction parameter related to the scattering length $a$,
$V({\bf r})$ is the random potential and the wave function is
normalized to the total number of atoms,
\begin{equation}\label{norm}
\int {d\bf r} |\Psi|^2 = N.
\end{equation}
We have written the equation in the BEC context and will use the
appropriate language throughout the paper (a brief translation
into the language of optics will be made at the end). The random
potential  can be generated optically (a laser speckle pattern,
\cite{hor}) or it can be due to some foreign (impurity) atoms,
with which the BEC interacts \cite{gav}. Eq. (\ref{G1}) is
applicable in any spatial dimension $d=1,2,3$. In the $2d$ case,
to which most of this paper is devoted, the condensate can move
only in the $(x, y)$-plane, whereas the motion in the
$z$-direction is frozen. For this geometry {\bf r} represents the
in-plane coordinate and $\triangle$ is the corresponding $2d$
Laplacian. The interaction parameter for the $2d$ problem is $g=
(8\pi)^{1/2}(\hbar^2 a/ma_z)$, where $a_z$ is the oscillator
length in the $z$-direction \cite{string}. We assume positive $g$,
i.e., repulsive interactions. Eq.(\ref{G1}) has to be solved with
the initial condition
\begin{equation}\label{in}
\Psi({\bf r}, t=0)=F({\bf r}),
\end{equation}
where F({\bf r}) represents the initial shape of the condensate,
i.e., the  equilibrium shape in the trap prior to its release. It
is assumed that at $t=0$ the trap is suddenly switched off and the
expansion, described by Eq. (\ref{G1}), starts.

We will be interested in strong nonlinearity in the sense that,
initially, almost the entire energy of the condensate is due to
interactions. In this case, and in the absence of the random
potential ($V({\bf r})=0$), the expansion is known to occur in two
stages \cite{kagan, castin, string}: a rapid "explosion", followed
by expansion linear in time. The explosive stage takes time of
order $t_0=1/\omega$, where $\omega$ is the frequency of the
parabolic trap, from which the BEC is released. We assume that the
trap is isotropic (for the $2d$ case the isotropy pertains, of
course, only to the $(x,y)$- plane but not to the z-direction).
During the time $t_0$ the condensate, roughly, doubles in size and
most of its interaction energy gets converted into the flow
energy. The second stage corresponds to times larger than $t_0$,
when the non-linearity becomes weak and the expansion is
accurately described by the linear Schr\"{o}dinger equation. This
observation suggests a simple, approximate approach to the problem
of expansion in the presence of disorder: If the disorder is
sufficiently weak (see the criterion below), then the first stage
of the expansion is dominated by strong nonlinearity, while the
disorder can be neglected. In the second stage the non-linearity
can be neglected, so that one is left with the linear
Schr\"{o}dinger equation, in the presence of disorder. This idea
has been proposed in \cite{san}, which appeared when the present
work was in progress (in \cite{san} the 1d case was considered,
whereas we are interested mostly in the 2d and 3d case). Thus, the
evolution of the BEC at the second stage of the expansion amounts
to a diffusion process in the weakly disordered potential.

 At the
end of the first stage of the expansion, i.e., at time of order
$t_0$, the condensate can be roughly described by a wave packet
\cite{kagan, castin, string}
\begin{equation}\label{packet}
\Psi(r, t_0)\simeq F(r)exp (ir^2/a_0^2),
\end{equation}
where $a_0 = (\hbar/m\omega)^{1/2}$ is the oscillator size of the
(in-plane) trap.  The function $F(r)$, on a qualitative level, can
be taken the same as the initial wave function in the trap:
indeed, during the time $t_0$ the condensate has only expanded by
a factor of 2 or so, preserving its shape of the inverted
parabola.  The essential thing that happens during the time $t_0$
is that the wave function acquires a phase, which accounts for the
flow energy. Since the velocity  of the condensate is proportional
to the gradient of the phase in Eq.(4), it follows that the
velocity is proportional to $r$, which is the main feature of the
ballistically expanding condensate.
 The wave function in
Eq.(\ref{packet}) serves as the initial condition for the second
stage of expansion, controlled by the linear Schr\"{o}dinger
equation, in the random potential $V({\bf r})$. To facilitate the
calculations below, we shall approximate the initial function by a
Gaussian, $F(r)=Aexp(-r^2/2R_0^2)$, where $A=(N/\pi R_0^2)^{1/2}$
is a normalization factor and $R_0$ characterizes the initial size
of the condensate. This replacement of the inverted parabola by a
Gaussian does not affect the evolution of the wave packet
(\ref{packet}) in any significant way.

A clear separation into the two stages can be made only if the
disorder is sufficiently weak \cite{san}. This means that during
the initial, "explosive" stage there is no scattering of the
condensate on the random potential, i.e., this initial stage is
ballistic. This requirement is satisfied if the transport mean
free path $l$ is larger than $2\pi/k$, where $k$ is the wave
number of the wave (particle) in question. The wave packet in
Eq.(\ref{packet}) contains a broad range of $k$'s, with a Gaussian
tail for $k>k_0$, where $k_0=R_0/a_0^2$. The value of $l$
separates the spatial Fourier components of the packet into two
groups. The components with $kl>>1$ will evolve by a diffusion
process, whereas those with $kl<1$ will be localized by disorder
and, thus, will stay in the vicinity of the initial location of
the packet. (In $2d$ also the large-$k$ components will eventually
get localized, but for those the localization length is
exponentially large so that there is a lot of room for the
diffusive propagation). In order to clearly observe the diffusive
behavior one must take $l\gg k_0^{-1}$. This is a rather weak
condition, because $k_0^{-1}$ is of the order of the healing
length $\xi$ which is the smallest macroscopic length in the
problem. (Note that we assume that the bottom of the trap, from
which the condensate was released, coincides with the average
value of the random potential; if the bottom were raised
sufficiently with respect to that value, then even the small-$k$
components would have enough energy to diffuse away.)
To ensure that the
great majority of the Fourier components would evolve by diffusion
we choose, somewhat arbitrarily, $l\approx R_0$, so that the
hierarchy of length scales is
\begin{equation}\label{scale}
\xi\ll a_0\ll R_0<l.
\end{equation}
Note that, since $R_0^2\simeq \mu/m\omega^2$ and $\xi^2\simeq
\hbar^2/m\mu$, the ratios $\xi/a_0$ and $a_0/R_0$ are of the same
order, $(\hbar \omega/\mu)^{1/2}$, where $\mu$ is the chemical
potential of the BEC before it was released from the trap.

Let us now present the calculation of the time evolution of a
packet, whose initial shape is defined in (\ref{packet}), with
$F(r)$ being a Gaussian, i.e.,
\begin{equation}\label{packet1}
\Psi(r,t=0)= (N/\pi R_0^2)^{1/2}exp(-r^2/2R_0^2)exp (ir^2/2a_0^2),
\end{equation}
where $t_0$ in the argument of $\Psi$ has been replaced by $0$.
This is because we are studying now the time evolution starting
from $t_0$ (and on time scale much larger than $t_0$). Decomposing
$\Psi(r,t=0)$ into Fourier components, $\tilde{\Psi}(k)$, we
obtain
\begin{equation}\label{FT}
\tilde{\Psi}(k)=(N/\pi R_0^2)^{1/2} b^2exp(-b^2k^2/2),
\end{equation}
where the complex length $b$ is defined as
\begin{equation}\label{b}
\frac{1}{b^2}=\frac{1}{R_0^2} -\frac{i}{a_0^2}.
\end{equation}

The quantity $\mid\tilde{\Psi}(k)\mid^2$ describes the density of
the particles in ${\bf k}$-space. More precisely, $dn(k)=
\mid\tilde{\Psi}(k)\mid^2d^2k$ is the number of particles  in the
interval $d^2k$, or in the interval $d^2v_k$ in the velocity space
$(v_k=\hbar k/m)$. The distribution $dn_k$, emerging from the
"explosion", will evolve by diffusion, due to scattering from the
random potential $V({\bf r})$. One must keep in mind that the
diffusion coefficient, $D(k)=\frac{1}{2}v(k)l(k)$ ( as well as the
mean free path $l$), depend on the velocity, so that the group
$dn(k)$ of particles, after time $t$, will  spread into a cloud
\begin{equation}\label{cloud}
dn_{{\bf k}} ({\bf r},t) =\mid\tilde{\Psi}(k)\mid^2
d^2k\frac{1}{4\pi D(k)t}exp(-\frac{r^2}{4D(k)t}).
\end{equation}
This expression has to be integrated  over {\bf k}, to obtain the
BEC shape, $\mid\Psi(r, t)\mid^2$, at time t. To perform this
integration we have to specify the dependence of $l$, or of the
mean transport time $\tau$, on $k$. For a Gaussian white noise
potential, $\tau$ is known to be inverse proportional to the
density of states $\nu (k)$ \cite{akk}. Since, in $2d$, $\nu
=const$, we have
$D(k)=\frac{1}{2}v^2(k)\tau=\frac{1}{2}\frac{\hbar^2k^2}{m^2}\tau$,
with $\tau$ being a constant specifying the strength of the random
potential. Integration over $k$ of the expression (\ref{cloud})
yields
\begin{equation}\label{shape}
\mid\Psi(r, t)\mid^2=\frac{N}{2\pi D_0 t}K_0 (\frac{r}{\sqrt{ D_0
t}}),
\end{equation}
where $K_0$ is the zeroth order modified Hankel function and $D_0$
is the diffusion coefficient corresponding to $k=k_0$, i.e., $D_0=
\frac{1}{2}v_0^2\tau=\frac{1}{2}\frac{\hbar^2k_0^2}{m^2}\tau$. The
mean square size of the condensate grows as $2D_0t$.

Eq.(\ref{shape}) describes the shape of the condensate in the
diffusion regime. It is quite remarkable that this shape depends
on the single quantity $D_0$. It is instructive to rewrite $D_0$
in terms of the chemical potential $\mu$ of the BEC in the trap,
prior to its expansion. Using $k_0^2=R_0^2/a_0^4=2m\mu /\hbar^2$,
we obtain $D_0=\mu\tau /m$. Thus, the shape (for a given $m$) is
determined by a single dimensionless parameter $\mu\tau /\hbar$.
The situation turns out to be analogous to a  disordered Fermi
system, where  $(E_F \tau /\hbar)\simeq k_F l$ is the only
parameter which determines the transport properties of a
disordered metal, with Fermi energy $E_F$ \cite{akk}. The
necessary condition for the diffusive regime described above is
that the parameter $(\mu\tau /\hbar)\gg1$. When the disorder
increases and $\mu\tau /\hbar$ approaches unity the diffusion is
inhibited and the BEC gets localized. (In $2d$, even for $(\mu\tau
/\hbar)\gg1$ spreading by diffusion will eventually stop and
localization is expected to set in. The localization length,
however, is exponentially large, $L_{loc}\sim exp(\mu\tau
/\hbar)$.) Instead of changing the disorder, one can tune the
interaction strength, i.e., the nonlinearity (for instance, using
a Feshbach resonance), and observe a crossover from localization
(weak interactions) to diffusion (strong interaction).

In the derivation of Eq.(\ref{shape}) we have ignored the fact
that the small-$k$ components of the  initial wave packet
(\ref{packet1}) cannot propagate by diffusion, as was already
discussed above. To account for this fact one should integrate the
expression (\ref{cloud}) with a lower cutoff, $k_{min}$. The
cutoff is determined by the condition $kl(k)\simeq 1$ , i.e.,
$k_{min}\simeq (m/\hbar\tau)^{1/2}$. Although it is not possible
then to derive a close analytic expression for $\mid\Psi(r,
t)\mid^2$, the qualitative picture is clear: most of the wave
packet will evolve according to Eq.(\ref{shape}), but a small
portion will remain localized near the origin. Furthermore, we
have  neglected  the weak localization effects, which influence
the large-$k$ components and renormalize (by a small amount) the
corresponding diffusion coefficient $D(k)$.

Our calculation can be extended to three dimensions. Eq.(9)
remains the same, with $d^3k$ instead of $d^2k$ and with
\begin{equation}\label{FT1}
\mid\tilde{\Psi}(k)\mid^2 =\frac{N}{(k_0\sqrt{\pi}
)^3}e^{-k^2/k_0^2}.
\end{equation}
The main difference with the two-dimensional case is that in $3d$
the density of states, $\nu(k)$, is proportional to k. This leads
to a linear dependence of the diffusion coefficient on $k$
(instead of the square dependence, in $2d$). The linear dependence
complicates integration over $k$, and we were not able to obtain a
close expression for  $\mid\Psi(r, t)\mid^2$. The essential thing,
however, is that the spread of $\mid\Psi(r, t)\mid^2$ is still
determined by the single quantity, $D_0=\frac{1}{3}v_0l=
\frac{1}{3}v_0^2\tau_0$, where $\tau_0\equiv \tau (k_0)$ (note
that, unlike in $2d$, $\tau$ now depends on $k$). Thus, again, the
behavior of the condensate is controlled by  the single parameter
$(\mu\tau_0 /\hbar)$, and diffusion is possible only if this
parameter is large. We will not dwell here on the diffusion regime
but rather  mention the interesting possibility of an
interaction-induced Anderson transition. The point is that, in
$3d$, there is a critical value, $c$, such that for $\mu\tau_0
/\hbar$ smaller than $c$ the BEC remains localized. Therefore, by
increasing the interaction strength, one can transfer the
condensate from the localized regime into the extended one.

Let us explain this crossover in more detail. For conceptual
simplicity it is useful not to create disorder in the trap itself
but only somewhat away. Then, in the absence of interactions $\mu$
would be small, of order $\hbar\omega$, and the condensate
released from the trap would not be able to penetrate the
disordered region (we assume that $\hbar\omega$ is smaller than
the mobility edge $E_c$). With the increase of interactions $\mu$,
as well as the size of the condensate $R_0$, increase and the wave
function, after the first (ballistic) stage of expansion, develops
oscillations (see Eq.(\ref{packet})). The number of oscillations
grows with $\mu$, so that for larger $\mu$ the wave packet in
Eq.(\ref{packet}) will contain more high energy components. The
components with energy larger than $E_c$ will be able to penetrate
the disordered region. Thus, under increase of interactions,
larger and larger portions of the condensate will be transferred
"over the mobility edge" and will diffuse away.

The specific dependence of $l$ (or $\tau$) on $k$ is determined by
the properties of the random potential. In our calculation a
Gaussian, white noise potential has been taken. For a correlated
potential the dependence of $l$ on $k$ will change (see, for
instance, \cite{apalkov, kuhn}). Let us return to the $2d$ case
and assume a Gaussian random potential, with zero mean, and with a
correlation function $<V({\bf r})V({\bf r'})>=V_0^2 exp(-\mid{\bf
r}-{\bf r'}\mid^2/R_c^2)$, where $V_0$ and $R_c$ characterize,
respectively, strength and range of the random potential. For this
case a close expression for the transport mean free path $l$ can
be obtained \cite {apalkov}:
\begin{equation}\label{mfp}
\frac{1}{l}=\frac{\pi}{k}(\frac{mV_0R_c}{\hbar^2})^2G(k^2R_c^2/2),
\end{equation}
where the function $G(x)$ is given in terms of the modified Bessel
functions as $G(x)=exp(-x)[I_0(x)-I_1(x)]$. For $kR_c\ll 1$ (the
white noise limit) $l$ is proportional to $k$. In the opposite
limit of a smooth random potential ($kR_c\gg 1$) $l$ is
proportional to $k^4$. $D(k)$ in (\ref{cloud}) becomes a
complicated function of $k$ and no closed expression for
$\mid\Psi(r, t)\mid^2$ can be obtained.  The detailed shape of the
expanding cloud is changed, as compared with the white noise case,
Eq.(10). The overall features, however, are preserved. For
instance, the mean square size of the cloud is still proportional
to $D_{eff}t$, where the effective diffusion coefficient depends
now on $R_c$ (in addition to $V_0$ and $k_0$). Thus,  the rough
picture of the expansion (but not the fine details) is universal
and depends on the single quantity, $D_{eff}$. This statement
pertains to short-range correlations, rapidly decaying beyond the
distance $R_c$ (the correlation radius). It has been emphasized
already in the original work of Anderson \cite{An} that long range
correlations in the randomness change the transport picture
completely. The "one-dimensional" speckles, designed in \cite{lye,
clement1, fort, clement2},  do exhibit long range correlations,
which has a significant effect on spreading of the $1d$ condensate
\cite{san}. It is not clear how long range correlations will
affect the diffusive spreading in 2 or 3 dimensions.

Let us briefly comment on the $1d$ case. There is some numerical
work on the subject \cite{ shep, kott} claiming that even weak
non-linearity (although still above a certain threshold) can lead
to complete delocalization. We should emphasize that weak
nonlinearity effects are not addressed in the present work. We
assumed strong nonlinearity, initially, and no nonlinearity after
the first, explosive stage of the expansion, when almost all
interaction energy had already been converted into flow. More
recent numerics \cite {akk1, modugno} concentrated on strong
nonlinearity, as appropriate for the experimental conditions
\cite{lye, clement1, fort,  clement2}, and identified various
regimes of $1d$ localization, depending on the degree of disorder.
In $1d$ the localization length, $L_{loc}$, is of the same order
as the mean free path $l$, so that there is no room for diffusion
and our results do not apply.  However, the classification of the
disorder as "weak" or "strong", depending on the value of the
parameter $k_0l\simeq (\mu\tau_0 /\hbar)$, remains meaningful. If
this parameter is large, then the details of the randomness, such
as the correlation radius of the random potential, are not
important (single parameter scaling \cite {A}). When this
parameter approaches unity, one enters the regime of strong
localization where details become important and single parameter
scaling is violated \cite {B}. For a correlated Gaussian
potential, $<V(x)V(x')>=V_0^2f(\mid x-x'\mid /R_c)$, the inverse
localization length, for a given wave number $k$, is of the order
of \cite{L} $L_{loc}^{-1}\simeq (mV_0/\hbar^2k)^2\tilde{f}(kR_c)$,
where $\tilde{f}(kR_c)$ is  the Fourier transform of some (short
range) correlation function $f(\mid x-x'\mid /R_c)$. Note that
$\tilde{f}(kR_c)$ has units of length, i.e., it can be written as
$R_cg(kR_c)$. For $kR_c\ll 1$, $g(kR_c)\approx 1$ (the white noise
limit). In this limit $L_{loc}$ is proportional to $k^2/V_0^2R_c$.
For $kR_c\gg 1$, $L_{loc}$ rapidly increases, due to the decrease
of the function $\tilde{f}(kR_c)$, so that the "weak disorder
condition", $kL_{loc}\gg 1$, becomes less restrictive.

 As has been mentioned at the beginning, Eq.(\ref{G1}) describes
also propagation in nonlinear optics, in the paraxial (scalar
wave) approximation. A monochromatic beam propagates in the
$z$-direction and spreads in the transverse, ($x,y$)-plane. To
translate Eq.(\ref{G1}) into the language of optics, one should
make the following replacements: $t\Longrightarrow z,
\hbar\Longrightarrow (c/\omega), m\Longrightarrow 1, V({\bf
r})\Longrightarrow (-\delta n(x,y)/n_0)$, where $\omega$ is the
frequency of the wave, $c$ is its speed (in the medium), $n_0$ is
the average refraction index, and $\delta n$ is its fluctuating
part (it is crucial that $\delta n$ does not depend on $z$).
Fluctuations of the refraction index in the transverse plane can
lead to "transverse localization" \cite {lag} of the wave. The
authors of \cite {segev} have recently observed this phenomenon,
as well as the interplay between nonlinearity and disorder. The
self-focusing (self-defocusing) in optics corresponds to
attractive (repulsive) interactions in the BEC. Thus, the results
presented in this paper applies to optics as well, for a
self-defocusing nonlinearity.

In conclusion, the diffusive spreading of a BEC, in the presence
of a random potential, has been studied. The spreading is
controlled by a single parameter, which can be conveniently
written as $(\mu\tau_{eff} /\hbar)$. Here $\tau_{eff}$ is some
effective  relaxation time which depends on the properties of the
random potential, as well as on $\mu$ (i.e., on $k_0=\sqrt{2m\mu}
/\hbar$). Diffusive spreading is possible only if this parameter
is large. There is a clear parallel with the single parameter
scaling in disordered electronic systems. There are also essential
differences between the two problems. In disordered conductors (at
low temperatures) only the electrons in the vicinity of the Fermi
energy participate in transport, whereas spreading of the BEC
involves a broad range of wave numbers $k$ (we considered only the
case when the center of mass of the condensate is fixed, i.e., the
condensate spreads but does not move as a whole). Strong repulsive
non-linearity, considered in this paper, has a large effect on the
BEC spreading and  serves as a strong "delocalizing factor".

Useful discussions with E. Akkermans, A. Minguzzi, M. Segev and J.
Steinhauer are gratefully acknowledged. The research was supported
by a grant from the Israel Science Foundation.


\begin{thebibliography}{99}


\bibitem{lye}J. E. Lye {\em et al.}, Phys. Rev. Lett. {\bf 95}, 070401
(2005).
\bibitem{clement1}D. Clement {\em et al.}, Phys. Rev. Lett. {\bf 95},
170409 (2005).
\bibitem{fort}C. Fort {\em et al.}, Phys. Rev. Lett. {\bf 95},
170410 (2005).
\bibitem{clement2} D. Clement {\em et al.}, New. J. Phys. {\bf 8},
165 (2006).
\bibitem{shulte} T. Schulte {\em et al.}, Phys. Rev. Lett. {\bf 95}, 170411
(2005).
\bibitem{segev} T. Schwartz, G. Bartal, S. Fishman and M. Segev,
Nature {\bf 446}, 52 (2007).
\bibitem{string} L. Pitaevskii and S. Stringari, "Bose-Einstein
Condensation", Clarendon Press, 2003.
\bibitem{hor} P. Horak, J.-Y. Courtois and G. Grynberg, Phys. Rev. A {\bf
58}, 3953 (1998).
\bibitem{gav}  U. Gavish and Y. Castin,   Phys. Rev. Lett. {\bf 95},
020401 (2005).
\bibitem{kagan}  Yu. Kagan, E. L. Surkov and G. V. Shlyapnikov,   Phys. Rev. A {\bf 54},
R1753 (1996).
\bibitem{castin} Y. Castin and R. Dum,   Phys. Rev. Lett. {\bf 77},
5315 (1996).
\bibitem{san} L. Sanchez-Palencia {\em et al.}, condmat/0612670.
\bibitem {akk} E. Akkermans and G. Montambaux, "Mesoscopic Physics
of Electrons and Photons", Cambridge University Press (2006).
\bibitem {apalkov} V. M. Apalkov, M. E. Raikh and B. Shapiro ,
 J. Opt. Soc. Am. B{\bf 21}, 132 (2004); and
 in "Anderson Localization and its Ramifications", Springer,
'Lecture Notes in Physics', ed. T. Brandes  and  S. Kettermann, p.
119, (2003).
\bibitem {kuhn} R. C. Kuhn {\em et al.}, Phys. Rev. Lett. {\bf 95}, 250403 (2005).
\bibitem {An} P. W. Anderson, Phys. Rev. {\bf 109}, 1492 (1958).
\bibitem {shep} D. L. Shepelyansky, Phys. Rev. Lett. {\bf 70},
1787 (1993).
\bibitem {kott} T. Kottos and M. Weiss, Phys. Rev. Lett. {\bf 93},
190604 (2004).
\bibitem {akk1} E. Akkermans, S. Ghosh and Z. Musslimani,
condmat/0610579 (to appear in Phys. Rev. A).
\bibitem {modugno} M. Modugno, Phys. Rev. A {\bf 73}, 013606
(2006).
\bibitem {A} E. Abrahams, P. W. Anderson, D. C. Licciardello and
T. V. Ramakrishnan, Phys. Rev. Lett. {\bf 42}, 673 (1979).
\bibitem {B} A. Cohen, Y. Roth and B. Shapiro, Phys. Rev. B {\bf
38}, 12125 (1988).
\bibitem {L} I. M. Lifshits, S. A. Gredeskul and L. A. Pastur,
"Introduction to the Theory of Disordered Systems", Wiley, 1988.
\bibitem {lag} H. De Raedt, A. Lagendijk and P. de Vries, Phys. Rev. Lett. {\bf 62}, 47
(1989).

\end{thebibliography}
\end{document}